\documentclass[prl,superscriptaddress,showpacs,amsmath,amssymb]{revtex4}
\usepackage{graphicx}
\usepackage{indentfirst}
\usepackage{psfrag}
\usepackage{epsfig}
\usepackage{amsmath}
\usepackage{amssymb}
\usepackage{bm}

\begin{document}
%%%%%%%%%%%%%%%%%% title page information %%%%%%%%%%%%%%%%%%
%{ \Large \bf Optical forces on small particles from partially coherent light}
%\\
%\\

%{\large Juan Miguel Au\~{n}\'{o}n  and Manuel Nieto-Vesperinas} 
%\\

\title{Optical forces on small particles from partially coherent light}

\author{Juan Miguel Au\~{n}\'{o}n  and Manuel Nieto-Vesperinas} 

\affiliation{Instituto de Ciencia de Materiales de Madrid, C.S.I.C., Campus de Cantoblanco, 28049, Madrid, Spain}

\email{mnieto@icmm.csic.es} %% email address is required



\begin{abstract}
We put forward a theory on the optical force exerted upon a dipolar particle by a stationary and ergodic partially coherent light field. We show through a rigorous analysis that the ensemble averaged electromagnetic force is given in terms of a partial gradient of the space variable diagonal elements of the coherence tensor. Further, by following this result we characterize  the conservative and non-conservative components of this force. In addition, we establish the propagation law for the optical force in terms of the coherence function of  light at a diffraction plane. This permits us to evaluate the effect of the degree of coherence on the force components  by using  the archetypical configuration of Young's two apertures diffraction pattern, so often employed to characterize coherence of waves.
\end{abstract}
%\ocis{(350.4855) Optical tweezers or optical manipulation; (260.2110) Electromagnetic optics; (030.1640) Coherence; %(030.6600) Statistical optics.} % REPLACE WITH CORRECT OCIS CODES FOR YOUR ARTICLE
\maketitle

%%%%%%%%%%%%%%%%%%%%%%% References %%%%%%%%%%%%%%%%%%%%%%%%%

%%%%%%%%%%%%%%%%%%%%%%%%%%  body  %%%%%%%%%%%%%%%%%%%%%%%%%%

\section{Introduction}

Since the works by Ashkin  \cite{Ashkin1970, Ashkin1986},
optical trapping and manipulation of particles became a tool of wide interest and use.  Optical tweezers of micrometric objects  are now  frequently employed in various areas of science, particularly in biology.

The mechanical action of light on particles is a consequence of its momentum \cite{jackson1998classical}. In this context some studies concerning momentum conservation laws have
been established both for deterministic \cite{ChaumetOL, Mansuripur04Radiation, Mansuripur07Radiation, Kemp05Abinitio} and
partially coherent wavefields \cite{kim2009momentum, Takeda}. However, in spite of the vaste research on coherence of light   (even in the context of inverse problems \cite{Carney1, Carney2, Garcia1, Garcia2})  and on optical trapping, no  study
on optical forces from fluctuating partially coherent fields has been developed yet, to our knowledge; apart from a couple of  works \cite{Li-GangWang,ChengliangZhao}  on Rayleigh particles in some particular focusing configurations. 

In the current knowledge on  photonic trapping \cite{nieto2004near, DholakiaGripped, Zelenina07Laser, QuidantNat07, albaladejo2009Scattering, Chaumet00Electromagnetic}, the illuminating light is assumed to be coherent, however, any field produced by a finite source is partially coherent \cite{MandelWolf, Perina, James1, James2, Dogariu}.  Addressing the partial coherence of light in optical manipulation of objects should become increasingly important as one enters in the nanoscale (or subwavelength region) in near-field studies \cite{nieto2004near}, and as a wider use is  done of 
 antenna-like sources with partial fluctuations as well as of the partial coherence induced in thermal sources.

In this paper we analyze in detail the different contributions to the optical forces exerted on a small particle by random stationary and ergodic partially coherent external fields \cite{MandelWolf}. We emphasize the influence of the degree of coherence on these forces. To this end, we address a system that,  since early studies, has been paradigmatic to observe the nature of light and matter waves, as well as to characterize the degree of coherence of wavefields \cite{MandelWolf}. This is the Young interference pattern from  two small apertures of an opaque screen. Concerning our study, we shall consider this configuration  as discussed in  the classical work by Thompson and Wolf \cite{thompson} that relates the observed visibility of the interference fringes with the estimated degree of coherence of the light  at those two apertures.

Hence, we shall establish a theory for the mean optical force on a dipolar particle, (understood as that whose electric and/or  magnetic polarizability is due to the corresponding first electric and/or magnetic Mie coefficient \cite{MNVOpex, GarciaEtxarri11Anistropic, MNVJOSA010}). This includes the limiting case of Rayleigh particles. Then we will study the dependence of this force gradient, scattering and curl components on the cross-spectral density of the fluctuating stationary wavefield. We  pay a special attention to the transition to a scalar theory  \cite{BornWolf} which characterizes most experiments in Fourier optics that do not measure or observe depolarization effects.  In this way, we present in Section 2 a rigorous formulation for the ensemble average of  the electromagnetic force; proving that this is given in terms of a partial gradient of the  space variable diagonal elements of the wavefield coherence tensor, (this tensor being electric and/or magnetic, depending on whether the particle responds to the electromagnetic wave with an electric and/or magnetic polarizability). Since this coherence function obeys the Zernike propagation law \cite{MandelWolf},  we prove that the optical force on a small particle from a partially coherent wavefield depends on the field coherence tensor, or cross-spectral density tensor, on a plane from which this wave has propagated.

Further, in Section 3 we consider the classical two-apertures configuration put forward in  Thompson and Wolf 's work \cite{thompson}. We perform calculations of the diffracted field, obtaining interesting interference pattern distributions for the different Cartesian components of the optical force exerted on a dipolar particle situated in a plane in the Fraunhofer region with respect to that of diffraction. Thus, we shall characterize the different magnitudes of these conservative and non-conservative force Cartesian components, both due to the diffraction process and to the degree of coherence of the light in the aperture plane, as well as steming from the particle polarizability.

\section{Averaged optical  force from a partially coherent wavefield}

We shall consider fluctuating time stationary and ergodic fields \cite{MandelWolf,BornWolf}.
For a single realization whose real electric and magnetic vectors
are $\mathbf{E}^{(r)}\left({\bf r},t\right)$ and $\mathbf{B}^{(r)}\left({\bf r},t\right)$,
respectively, at a space point ${\bf r}$ and time $t$, the frequency
decomposition is \cite{MandelWolf} 
\begin{eqnarray}
\mathbf{E}^{(r)}\left({\bf r},t\right) & = & \int_{-\infty}^{\infty}\mathbf{\tilde{E}}^{(r)}\left(\mathbf{r},\omega\right)e^{-i\omega t}d\omega,\label{EFT}\\
\mathbf{B}^{(r)}\left({\bf r},t\right) & = & \int_{-\infty}^{\infty}\tilde{\mathbf{B}}^{(r)}\left(\mathbf{r},\omega\right)e^{-i\omega t}d\omega\label{BFT}
\end{eqnarray}
 The corresponding complex analytic signals are \cite{MandelWolf,BornWolf}:
\begin{eqnarray}
\mathbf{E}\left({\bf r},t\right) & = & \int_{-\infty}^{\infty}\mathbf{\tilde{E}}\left(\mathbf{r},\omega\right)e^{-i\omega t}d\omega,\label{Eanal}\\
\mathbf{B}\left({\bf r},t\right) & = & \int_{-\infty}^{\infty}\mathbf{\tilde{B}}\left(\mathbf{r},\omega\right)e^{-i\omega t}d\omega,\label{Banal}
\end{eqnarray}
 where those Fourier integrals should be considered in the sense of
distribution theory. In addition, we have 
\begin{equation}
\begin{array}{cccc}
\mathbf{\tilde{E}}\left(\mathbf{r},\omega\right) & = & \mathbf{\tilde{E}}^{(r)}\left(\mathbf{r},\omega\right) & ,\omega\geq0\\
 & = & 0 & ,\omega<0
\end{array},\label{Eespectra}
\end{equation}

\begin{flushleft}
\begin{equation}
\begin{array}{cccc}
\mathbf{\tilde{B}}\left(\mathbf{r},\omega\right) & = & \mathbf{\tilde{B}}^{(r)}\left(\mathbf{r},\omega\right) & ,\omega\geq0\\
 & = & 0 & ,\omega<0
\end{array},\label{Bespectra}
\end{equation}
 
\begin{eqnarray}
\mathbf{E}\left({\bf r},t\right) & = & \frac{1}{2}\left[\mathbf{E}^{(r)}\left(\mathbf{r},t\right)+i\mathbf{E}^{(i)}\left(\mathbf{r},t\right)\right],\label{Eysumma}\\
\mathbf{B}\left({\bf r},t\right) & = & \frac{1}{2}\left[\mathbf{B}^{(r)}\left(\mathbf{r},t\right)+i\mathbf{B}^{(i)}\left(\mathbf{r},t\right)\right].\label{Bysuma}
\end{eqnarray}

\par\end{flushleft}

The superscripts $(r)$ and $(i)$ denote the real and imaginary parts,
respectively. For each Cartesian component of these electric and magnetic
vectors, they form a Hilbert transform pair in the $t$ variable \cite{MandelWolf,BornWolf}.

We shall now calculate the {\it ensemble average} of the force exerted by
the random field on a dipolar particle, (understood in the sense mentioned in Section 1), over its different realizations:
\begin{equation}
\left\langle {\bf F}({\bf r},t)\right\rangle =\left\langle \left({\bf p}^{(r)}({\bf r},t)\cdot\nabla\right){\bf E}^{(r)}({\bf r},t)+\frac{1}{c}\frac{\partial{\bf p}^{(r)}({\bf r},t)}{\partial t}\times{\bf B}^{(r)}({\bf r},t)\right\rangle ,\label{F}
\end{equation}
 ${\bf p}^{(r)}$ is the real part of the dipole moment induced by
the fluctuating incident wave on the particle. If $\alpha_{e}$ denotes
the particle electric polarizability, one has that 
\begin{equation}
\mathbf{p}(\mathbf{r},t)=\alpha_{e}\mathbf{E}(\mathbf{r},t)\label{p,E}
\end{equation}

Denoting: $\partial p/\partial t=\dot{p}$, and due to a well-known
property of the derivative of Hilbert transforms \cite{MandelWolf},
the real and imaginary parts of each Cartesian component of $\dot{{\bf p}}$
are Hilbert transforms of each other in $t$.

Let us evaluate the fist term of Eq. (\ref{F}). Taking Eqs. (\ref{Eanal}), (\ref{Banal}), (\ref{Eysumma}), (\ref{Bysuma}) and (\ref{p,E})
into account, omitting the explicit ${\bf r}$, $t$ dependence in
the forthcoming notation; this is (cf. \cite{MandelWolf,BornWolf,Perina}):
\begin{eqnarray}
& & \left\langle p_{j}^{(r)}\partial_{j}E_{i}^{(r)}\right\rangle \nonumber \\
 & = & \left\langle \left(p_{j}+p_{j}^{*}\right)\partial_{j}\left(E_{i}+E_{i}^{*}\right)\right\rangle \nonumber \\
 & = & \underset{\text{T}\rightarrow\infty}{\text{Lim}}\frac{1}{2T}\int_{-T}^{T}dt\left[\int_{-\infty}^{\infty}e^{-i(\omega_{1}+\omega_{2})t}\left\langle \tilde{p}_{j}({\bf r},\omega_{1})\partial_{j}\tilde{E}_{i}(\mathbf{r},\omega_{2})\right\rangle d\omega_{1}d\omega_{2}\right.\nonumber \\
 & + & \int_{-\infty}^{\infty}e^{-i\left(\omega_{1}-\omega_{2}\right)t}\left\langle \tilde{p_{j}}(\mathbf{r},\omega_{1})\partial_{j}\tilde{E_{i}}^{*}(\mathbf{r},\omega_{2})\right\rangle d\omega_{1}d\omega_{2}\nonumber \\
 & + & \int_{-\infty}^{\infty}e^{-i\left(-\omega_{1}+\omega_{2}\right)t}\left\langle \tilde{p_{j}}^{*}(\mathbf{r},\omega_{1})\partial_{j}\tilde{E_{i}}(\mathbf{r},\omega_{2})\right\rangle d\omega_{1}d\omega_{2}\nonumber \\
 & + & \left.\int_{-\infty}^{\infty}e^{+i\left(\omega_{1}+\omega_{2}\right)t}\left\langle \tilde{p_{j}}^{*}(\mathbf{r},\omega_{1})\partial_{j}\tilde{E_{i}}^{*}(\mathbf{r},\omega_{2})\right\rangle d\omega_{1}d\omega_{2}\right],
\end{eqnarray}
 Where $i,j=1,2,3$ and Einstein's convention of omitting the sum
symbol $\sum_{j=1}^{3}$ on the repeated index $j$ has been used.
Using the properties of Dirac delta distribution: $2\pi \delta(\omega)=\int_{-\infty}^{\infty}e^{-i\omega t}dt$
and $\delta(\omega)=\delta(-\omega)$, the previous equation reduces
to

\begin{eqnarray}
\left\langle p_{j}^{(r)}\partial_{j}E_{i}^{(r)}\right\rangle  & = & 2\pi\left[\int_{-\infty}^{\infty}g_{i}^{(p,E)}\left(\mathbf{r},\omega_{1},\omega_{2}\right)\delta\left(\omega_{1}+\omega_{2}\right)d\omega_{1}d\omega_{2}\right.\nonumber \\
 & + & \int_{-\infty}^{\infty}g_{i}^{(p,E^{*})}\left(\mathbf{r},\omega_{1},\omega_{2}\right)\delta\left(\omega_{1}-\omega_{2}\right)d\omega_{1}d\omega_{2}\nonumber \\
 & + & \int_{-\infty}^{\infty}g_{i}^{(p^{*},E)}\left(\mathbf{r},\omega_{1},\omega_{2}\right)\delta\left(\omega_{1}-\omega_{2}\right)d\omega_{1}d\omega_{2}\nonumber \\
 & + & \left.\int_{-\infty}^{\infty}g_{i}^{(p^{*},E^{*})}\left(\mathbf{r},\omega_{1},\omega_{2}\right)\delta\left(\omega_{1}+\omega_{2}\right)d\omega_{1}d\omega_{2}\right],\label{F1}
\end{eqnarray}
 where the cross-spectral function $g_{i}^{(U,V)}({\bf r},\omega_{1},\omega_{2})$
is 
\begin{equation}
g_{i}^{(U,V)}({\bf r},\omega_{1},\omega_{2})=\underset{\text{T}\rightarrow\infty}{\text{Lim}}\frac{1}{2T}\left\langle \tilde{U}_{j}({\bf r},\omega_{1})\partial_{j}\tilde{V}_{i}(\mathbf{r},\omega_{2})\right\rangle ,\label{g}
\end{equation}
$\tilde{U}_{j}(\mathbf{r},\omega)$ and $\tilde{V}_{i}(\mathbf{r},\omega)$ $\left(i,j=1,2,3\right)$
being the spectra of two analytic signal Cartesian components.

On performing the $\omega_{2}$ integration in Eq. (\ref{F1}) and
taking into account that due to Eqs. (\ref{Eanal}) - (\ref{Bespectra})
, and to Eq. (\ref{g}), one has that 
\begin{equation}
g_{i}^{(U,V)}({\bf r},\omega_{1},-\omega_{1})=0,\label{zero terms1}
\end{equation}
 only the second and third terms of Eq. (\ref{F1}) are different
from zero. Thus finally 
\begin{eqnarray}
\left\langle p_{j}^{(r)}\partial_{j}E_{i}^{(r)}\right\rangle  & = & 4\pi\Re\int_{-\infty}^{\infty}\underset{\text{T}\rightarrow\infty}{\text{Lim}}\frac{1}{2T}\left\langle \tilde{p_{j}}(\mathbf{r},\omega_{1})\partial_{j}\tilde{E_{i}}^{*}(\mathbf{r},\omega_{1})\right\rangle d\omega_{1}\nonumber \\
 & = & 4\pi\Re\int_{-\infty}^{\infty}g_{i}^{(p^{*},E)}\left({\bf r},\omega_{1},\omega_{1}\right)d\omega_{1},\label{F1anal}
\end{eqnarray}
 where $\Re$ denotes the real part.

On the other hand, writing as $\epsilon_{ijk}$ the antisymmetric
Levi \textendash{} Civita tensor, ($i,j,k=1,2,3$); the second term
of Eq. (\ref{F}) is: 
\begin{eqnarray}
\frac{1}{c}\epsilon_{ijk}\left\langle \dot{p}_{j}^{(r)}B_{k}^{(r)}\right\rangle  & = & \frac{1}{c}\epsilon_{ijk}\left\langle (\dot{p}_{j}+\dot{p}_{j}^{\ast})(B_{k}+B_{k}^{\ast})\right\rangle 
\end{eqnarray}
 
\begin{eqnarray}
 &  & \frac{1}{c}\epsilon_{ijk}\left\langle (\dot{p}_{j}+\dot{p}_{j}^{\ast})(B_{k}+B_{k}^{\ast})\right\rangle \nonumber \\
 & = & \frac{\epsilon_{ijk}}{c}\underset{\text{T}\rightarrow\infty}{\text{Lim}}\frac{1}{2T}\int_{-T}^{T}dt\left[\int_{-\infty}^{\infty}(-i\omega_{1})e^{-i(\omega_{1}+\omega_{2})t}\left\langle \tilde{p}_{j}({\bf r},\omega_{1})\tilde{B}_{k}(\mathbf{r},\omega_{2})\right\rangle d\omega_{1}d\omega_{2}\right.\nonumber \\
 & + & \int_{-\infty}^{\infty}(-i\omega_{1})e^{-i\left(\omega_{1}-\omega_{2}\right)t}\left\langle \tilde{p_{j}}(\mathbf{r},\omega_{1})\tilde{B_{k}}^{*}(\mathbf{r},\omega_{2})\right\rangle d\omega_{1}d\omega_{2}\nonumber \\
 & + & \int_{-\infty}^{\infty}i\omega_{1}e^{-i\left(-\omega_{1}+\omega_{2}\right)t}\left\langle \tilde{p_{j}}^{*}(\mathbf{r},\omega_{1})\tilde{B_{k}}(\mathbf{r},\omega_{2})\right\rangle d\omega_{1}d\omega_{2}\nonumber \\
 & + & \left.\int_{-\infty}^{\infty}i\omega_{1}e^{+i\left(\omega_{1}+\omega_{2}\right)t}\left\langle \tilde{p_{j}}^{*}(\mathbf{r},\omega_{1})\tilde{B_{k}}^{*}(\mathbf{r},\omega_{2})\right\rangle d\omega_{1}d\omega_{2}\right]
\end{eqnarray}
 Or in a more compact form 
\begin{eqnarray}
 &  & \frac{1}{c}\epsilon_{ijk}\left\langle (\dot{p}_{j}+\dot{p}_{j}^{\ast})(B_{k}+B_{k}^{\ast})\right\rangle \nonumber \\
 & = & \frac{2\pi}{c}\epsilon_{ijk}\left[\int_{-\infty}^{\infty}(-i\omega_{1})W_{jk}^{(p,B)}({\bf r},\omega_{1},\omega_{2})\delta(\omega_{1}+\omega_{2})d\omega_{1}d\omega_{2}\right.\nonumber \\
 & + & \int_{-\infty}^{\infty}(-i\omega_{1})W_{jk}^{(p,B^{\ast})}({\bf r},\omega_{1},\omega_{2})\delta(\omega_{1}-\omega_{2})d\omega_{1}d\omega_{2}\nonumber \\
 & + & \int_{-\infty}^{\infty}i\omega_{1}W_{jk}^{(p^{\ast},B)}({\bf r},\omega_{1},\omega_{2})\delta(\omega_{1}-\omega_{2})d\omega_{1}d\omega_{2}\nonumber \\
 & + & \left.\int_{-\infty}^{\infty}i\omega_{1}W_{jk}^{(p^{\ast},B^{\ast})}({\bf r},\omega_{1},\omega_{2})\delta(\omega_{1}+\omega_{2})d\omega_{1}d\omega_{2}\right],\label{F2}
\end{eqnarray}
 where the cross-frequency density tensor $W_{jk}^{(U,V)}({\bf r},\omega_{1},\omega_{2})$ is 
\begin{equation}
W_{jk}^{(U,V)}({\bf r},\omega_{1},\omega_{2})=\underset{\text{T}\rightarrow\infty}{\text{Lim}}\frac{1}{2T}\left\langle \tilde{U}_{j}({\bf r},\omega_{1})\tilde{V}_{k}(\mathbf{r},\omega_{2})\right\rangle , \label{crosssfreq}
\end{equation}

On performing the $\omega_{2}$ integration in Eq. (\ref{F2}) and
taking into account that due to Eqs. (\ref{Banal}) - (\ref{Bespectra}),
one has that 
\begin{equation}
W_{jk}^{(U,V)}({\bf r},\omega_{1},-\omega_{1})=0\label{zero terms}
\end{equation}
 only the second and third terms of Eq. (\ref{F2}) remain different
from zero.

Now, since ${\bf B}=c/i\omega\nabla\times{\bf E}$, i.e. $-i\omega B_{k}^{\ast}=c\epsilon_{klm}\partial_{l}E_{m}^{\ast}$,
and taking into account that $\epsilon_{ijk}\epsilon_{klm}=\delta_{il}\delta_{jm}-\delta_{im}\delta_{jl}$,
$\delta_{il}$ being the Kronecker delta unit tensor, Eq. (\ref{F2})
becomes 
\begin{eqnarray}
 &&\frac{1}{c}\epsilon_{ijk}\left\langle \dot{p}_{j}^{(r)}B_{k}^{(r)}\right\rangle \nonumber \\ &=&4\pi\Re\int_{-\infty}^{\infty}\underset{\text{T}\rightarrow\infty}{\text{Lim}}\frac{1}{2T}\left[\left\langle \tilde{p_{j}}(\mathbf{r},\omega_{1})\partial_{i}\tilde{E_{j}}^{*}(\mathbf{r},\omega_{1})\right\rangle \right.\nonumber\\
 &-&\left.\left\langle \tilde{p_{j}}(\mathbf{r},\omega_{1})\partial_{j}\tilde{E_{i}}^{*}(\mathbf{r},\omega_{1})\right\rangle \right]d\omega_{1}. \label{F2anal}
\end{eqnarray}
We introduce the electric field {\it  coherence tensor} \cite{MandelWolf, Perina} ${\cal E}_{jk} ({\bf r}, {\bf r}',\tau)=\left\langle {E_{j}}(\mathbf{r},t){E_{k}}^{*}(\mathbf{r}',t+\tau)\right\rangle$ expressed as
\begin{equation}
{\cal E}_{jk} ({\bf r}, {\bf r}',\tau)=\int_{-\infty}^{-\infty}\tilde{\cal E}_{jk} ({\bf r}, {\bf r}',\omega) e^{-i\omega \tau}d\omega, \label{tensorEt}
\end{equation}

$\tilde{\cal E}_{jk} ({\bf r}, {\bf r}',\omega)$ being the electric field {\it cross-spectral density tensor} defined as
\begin{equation}
\tilde{\cal E}_{jk} ({\bf r}, {\bf r}',\omega)=\underset{\text{T}\rightarrow\infty}{\text{Lim}}\frac{1}{2T}\left\langle \tilde{E}_{j}({\bf r},\omega_{1})\tilde{E}_{k}(\mathbf{r}',\omega_{2})\right\rangle . \label{crossspect}
\end{equation}
 On introducing Eqs. (\ref{F1anal}) and (\ref{F2anal}) into Eq.
(\ref{F}), and  taking Eqs. (\ref{p,E}), (\ref{tensorEt}) and (\ref{crossspect})  into acount,  one finally obtains for the averaged force acting on a dipolar particle \cite{ChaumetOL}, the following expression in terms of the analytic signal associated to the random field 
\begin{eqnarray}
\left\langle F_{i}({\bf r},t)\right\rangle  & = &    4\pi\Re\int_{-\infty}^{\infty}\underset{\text{T}\rightarrow\infty}{\text{Lim}}\frac{1}{2T}\left\langle \tilde{p_{j}}(\mathbf{r},\omega)\partial_{i}\tilde{E_{j}}^{*}(\mathbf{r},\omega)\right\rangle d\omega\nonumber \\
 & = & 4\pi\Re\int_{-\infty}^{\infty}\alpha_{e}\partial_{i}^{(*)}\underset{T\rightarrow\infty}{Lim}\frac{1}{2T}\left\langle \tilde{E_{j}}(\mathbf{r},\omega)\tilde{E_{j}}^{*}(\mathbf{r},\omega)\right\rangle d\omega\nonumber \\
& = & 4\pi\Re\int_{-\infty}^{\infty}\alpha_{e}\partial_{i}^{(*)} Tr \tilde{\cal E}_{jk} ({\bf r}, {\bf r},\omega) d\omega, \label{ftotW}
\end{eqnarray}
 where we have substituted $\omega_{1}$ by $\omega$. The symbol  {\it Tr} denotes the {\it trace}. On the other hand, $\partial_{i}^{(*)}$ means that the derivative with respect to the $ ith$ component of ${\bf r}$ is made on the complex-conjugated component $E_{j}^{*}$. 

Eq. (\ref{ftotW}) may also be expressed in terms of the coherence tensor as
\begin{eqnarray}
\left\langle F_{i}({\bf r},t)\right\rangle
  & = & 4\pi\Re\left\{ \left\langle p_{j}(\mathbf{r},t)\partial_{i}E_{j}^{*}(\mathbf{r},t)\right\rangle \right\}= \nonumber 
% & = & 
4\pi\Re\left\{ \alpha_{e}\partial_{i}^{(*)}\left\langle E_{j}(\mathbf{r},t)E_{j}^{*}(\mathbf{r},t)\right\rangle \right\} \nonumber \\
 & = & 4\pi\Re\left\{ \alpha_{e}\partial_{i}^{(*)} Tr {\cal E}_{jk}({\bf r},{\bf r},0) \right\} . \label{ftot1}
\end{eqnarray}
 In Eq. (\ref{ftot1}) we have recalled that acording to  (\ref{tensorEt}) one has that ${\cal E}_{jk} ({\bf r}, {\bf r},0) = \int_{-\infty}^{\infty} \tilde{\cal E}_{jk} ({\bf r}, {\bf r},\omega)d\omega$.   
On introducing the mean force  Fourier components $\left\langle \tilde{ F}_{i}({\bf r},\omega)\right\rangle$ as
\begin{equation}
\left\langle  F_{i}({\bf r},t)\right\rangle=2\pi \int_{-\infty}^{\infty}\left\langle \tilde{F}_{i}({\bf r},\omega)\right\rangle d\omega, \label{Fomega1}
\end{equation}
 we see that according to Eq. (\ref{ftotW})  we may express them as:
\begin{equation}
\left\langle  \tilde{F}_{i}({\bf r}, \omega)\right\rangle  
=2\Re [\alpha_{e}\partial_{i}^{(*)} Tr \tilde{\cal E}_{jk} ({\bf r}, {\bf r},\omega)]   . \label{Fomega2}
\end{equation}

The above calculations
also lead to the conclusion that , being ${\bf p}$, $\dot{{\bf p}}$,
${\bf E}$ and ${\bf B}$ analytic signals of $t$, one also has 
that 
\begin{eqnarray}
\left\langle \dot{p}_{j}^{(r)}B_{k}^{(r)}\right\rangle  & = & 2\Re\left\langle \dot{p}_{j}B_{k}^{*}\right\rangle ,\nonumber \\
\left\langle p_{j}^{(r)}\partial_{j}E_{i}^{(r)}\right\rangle  & = & 2\Re\left\langle E_{j}\partial_{j}E_{i}^{*}\right\rangle .\label{cross}
\end{eqnarray}
 Notice that introducing Eqs. (\ref{cross}) into Eq. (\ref{F}),
one obtains again (\ref{ftot1}).

Eq. (\ref{ftot1}) shows that the mean  force is linked to the coherence tensor of the field. This latter quantity fulfills the Helmholtz
equation whose integral representation leads to well known propagation laws in
coherence theory, like the Zernike  law and the Van Cittert-Zernike theorem, or to the dependence of the intensity on the degree of coherence of the wavefield in the primary
or secondary source surface that emits it \cite{MandelWolf,BornWolf,Perina}.
Hence all these phenomena have consequences for the averaged force.

\subsection{Conservative and non-conservative components of the averaged optical force. The case of magnetodielectric particles}
It is well-known that the time averaged force from coherent fields may be expressed as the sum of three parts, {\it one conservative and two non-conservative}, (cf.  \cite{albaladejo2009Scattering, MNVOpex}), namely,  a {\it gradient,
a scattering and a curl of a electric spin density}. Similarly, the
force spectral components given by Eq. (\ref{Fomega2}) lead to 
\begin{eqnarray}
\left\langle \mathbf{\tilde{F}}(\mathbf{r},t)\right\rangle  & = & 2\pi\Re\alpha_{e}\int_{-\infty}^{+\infty}\underset{\text{T}\rightarrow\infty}{\text{Lim}}\frac{1}{2T}\left\langle \nabla\left|\mathbf{\tilde{E}}(\mathbf{r},\omega)\right|^{2}\right\rangle d\omega\nonumber \\
 & + & 4\pi\Im\alpha_{e}\Re\left\{ \int_{-\infty}^{+\infty}\underset{\text{T}\rightarrow\infty}{\text{Lim}}\frac{1}{2T}\left\langle k\mathbf{\tilde{E}}(\mathbf{r},\omega)\times\mathbf{\tilde{B}}^{*}(\mathbf{r},\omega)\right\rangle d\omega\right\} \nonumber \\
 & + & 4\pi\Im\alpha_{e}\Im\left\{ \int_{-\infty}^{+\infty}\underset{\text{T}\rightarrow\infty}{\text{Lim}}\frac{1}{2T}\left\langle \left(\mathbf{\tilde{E}^{*}}(\mathbf{r},\omega)\cdot\nabla\right)\mathbf{\tilde{E}}(\mathbf{r},\omega)\right\rangle d\omega\right\} ,\label{3.4.20}
\end{eqnarray}
 where $\Im$ denotes the imaginary part.  In Eq. (\ref{3.4.20}) the first term represents the conservative or {\it gradient} force, whereas the second and third terms correspond to the non-conservative {\it scattering component}, or {\it radiation pressure}, and  to the {\it curl} force, respectively. Likewise, one may write the same decomposition for the averaged
force spectral components in $\omega-$space, [cf. Eq. (\ref{Fomega2})]: 
\begin{eqnarray}
\left\langle \tilde{\mathbf{F}}\left(\mathbf{r},\omega\right)\right\rangle &=&\Re\alpha_{e}\nabla\left\langle \left|\tilde{\mathbf{E}}\left(\mathbf{r},\omega\right)\right|^{2}\right\rangle +2k\Im\alpha_{e}\Re\left\{ \left\langle \tilde{\mathbf{E}}\left(\mathbf{r},\omega\right)\times\tilde{\mathbf{B}}^{*}\left(\mathbf{r},\omega\right)\right\rangle \right\} \nonumber \\
&+&2\Im\alpha_{e}\Im\left\{ \left\langle \left(\tilde{\mathbf{E}}^{*}\left(\mathbf{r},\omega\right)\cdot\nabla\right)\tilde{\mathbf{E}}\left(\mathbf{r},\omega\right)\right\rangle \right\}  . \label{graddecomp}
\end{eqnarray}

\noindent It should be remarked that if the particle is magnetodielectric, namely, if  additionally it has a magnetic
polarizability $\alpha_{m}$ \cite{MNVOpex}, then in a similar way as for 
Eq. \eqref{graddecomp} one obtains for the averaged force on the
particle due to the magnetic field $\bar{F}^{m}_{i}({\bf r},t)=2\Re\left\{ \alpha_{m}\partial_{i}^{(*)} Tr {\cal B}_{jk}({\bf r},{\bf r},0) \right\}$, $ {\cal B}_{jk}({\bf r},{\bf r}',\tau)=\left\langle B_{j}(\mathbf{r},t)B_{k}^{*}(\mathbf{r},t+\tau)\right\rangle$:
\begin{eqnarray}
\left\langle \tilde{\mathbf{F}}^{m}\left(\mathbf{r},\omega\right)\right\rangle &=&\Re\alpha_{m}\nabla\left\langle \left|\tilde{\mathbf{B}}\left(\mathbf{r},\omega\right)\right|^{2}\right\rangle +2k\Im\alpha_{m}\Re\left\{ \left\langle \tilde{\mathbf{E}}\left(\mathbf{r},\omega\right)\times\tilde{\mathbf{B}}^{*}\left(\mathbf{r},\omega\right)\right\rangle \right\} \nonumber \\
&+&2\Im\alpha_{m}\Im\left\{ \left\langle \left(\tilde{\mathbf{B}}^{*}\left(\mathbf{r},\omega\right)\cdot\nabla\right)\tilde{\mathbf{B}}\left(\mathbf{r},\omega\right)\right\rangle \right\}  .\label{magnetic}
\end{eqnarray}

And for the mean force due to the interaction between the electric and magnetic dipole induced in the particle $\left\langle {F}_{i}^{e-m}({\bf r},t) \right\rangle=-(8/3)k^{4}\Re\left\{ \alpha_{e}\alpha_{m}^{*}\epsilon_{ijk}{\cal G}_{jk}({\bf r},{\bf r},0)\right\}$,  ${\cal G}_{jk}({\bf r},{\bf r}',\tau)=\left\langle E_{j}(\mathbf{r},t)B_{k}^{*}(\mathbf{r},t+\tau)\right\rangle $ \cite{MNVOpex}: 
\begin{eqnarray}
\left\langle \mathbf{\tilde{F}}^{e-m}\right\rangle  & = & -\frac{4k^{4}}{3}\left\{ \Re(\alpha_{e}\alpha_{m}^{*})\Re\left\langle \mathbf{\tilde{E}}\times\mathbf{\tilde{B}}^{*}\right\rangle - \Im(\alpha_{e}
\alpha_{m}^{*})\Im\left\langle \mathbf{\tilde{E}}\times\mathbf{\tilde{B}}^{*}\right\rangle \right\} \nonumber \\
 & = & -\frac{4k^{4}}{3}\Re(\alpha_{e}\alpha_{m}^{*})\Re \left\langle \mathbf{\tilde{E}}\times\mathbf{\tilde{B}}^{*}\right\rangle \nonumber \\
 & + & \frac{4k^{3}}{3}\Im(\alpha_{e}\alpha_{m}^{*})\left[\frac{1}{2}\nabla\left\langle \left|\tilde{\mathbf{E}}\right|^{2}\right\rangle -\Re\left\langle (\mathbf{\tilde{E}}^{*}\cdot\nabla)\mathbf{\tilde{E}}\right\rangle \right].\label{em}
\end{eqnarray}
For the sake of brevity, we have omitted in the notation of Eq.  (\ref{em}) the arguments ${\bf r}$ and $\omega$ of the analytic
signal spectral vectors $\mathbf{\tilde{E}}$ and $\mathbf{\tilde{B}}$. In this paper, we study
the mean force on a particle with electric polarizability $\alpha_{e}$ only, [cf. Eq. \eqref{graddecomp}].

\subsection{Dependence of the averaged optical force of propagated fields on the coherence at  a diffraction plane. A Young interference configuration}

To illustrate the above with a simple example, let us consider a wavefield
whose frequency components may be described by a scalar function $U({\bf r},\omega)$. (This space-frequency description may also apply to a quasimonochromatic
field, harmonically vibrating as $\exp(-i\bar{\omega}t)$ around a mean frequency $\bar{\omega}$). In this case ${\bf \tilde{E}}({\bf r},\omega)=U({\bf r},\omega){\bf e}(\omega)$,
(cf. Section 8.4 of \cite{BornWolf}). This a  common situation in Fourier optics \cite{goodmam1996}. The vector ${\bf e}(\omega)$ is real  (linear polarization). On introducing  $W({\bf r}_{1},{\bf r}_{2},\omega)=\left\langle U^{*}({\bf r}_{1},\omega)U({\bf r}_{2},\omega)\right\rangle $  as the {\it cross-spectral density} of  $U({\bf r},\omega)$ \cite{MandelWolf},   writing $\partial_{i}^{(*)}W({\bf r},{\bf r},\omega)=\left\langle U({\bf r},\omega)\partial_{i}U^{*}({\bf r},\omega)\right\rangle $
and taking real and imaginary parts in Eq. (\ref{Fomega2}) one obtains
\begin{eqnarray}
\left\langle \tilde{F}_{i}\left(\mathbf{r},\omega\right)\right\rangle &=&   
2|{\bf e}(\omega)|^{2}\Re\left\{ \alpha_{e}\left\langle \partial_{i}U^{*}({\bf r},\omega)U({\bf r},\omega)\right\rangle \right\} \nonumber \\
&=& 2 |{\bf e}(\omega)|^{2}[\Re\alpha_{e}\Re\left\langle U({\bf r},\omega)\partial_{i}U^{*}
({\bf r},\omega)\right\rangle - \Im\alpha_{e} \Im\left\langle U({\bf r},\omega)\partial_{i}U^{*}({\bf r},\omega)\right\rangle].  \label{Fescal}
\end{eqnarray}
The first term of Eq. (\ref{Fescal}), is the {\it mean gradient force}, which is expressed as [see also the first term of Eq. (\ref{graddecomp})]:
 
\begin{eqnarray}
\left\langle \tilde{F}_{i}^{grad}\left(\mathbf{r},\omega\right)\right\rangle  & = & |{\bf e}(\omega)|^{2}\Re\alpha_{e}\partial_{i}W({\bf r},{\bf r},\omega)\nonumber \\
 & = & |{\bf e}(\omega)|^{2}\Re\alpha_{e}\partial_{i}\left\langle |U({\bf r},\omega)|^{2}\right\rangle .\label{Fgrad}
\end{eqnarray}
The second term of Eq. (\ref{Fescal}) is proportional to the {\it mean energy flow spectral density} $\left\langle \cal{S} \right\rangle$ associated to the scalar wavefunction $U({\bf r},\omega)$ \cite{nietolibro}:
 \begin{eqnarray}
\left\langle {\cal S}_{i}({\bf r},\omega) \right\rangle =-\frac{1}{k}\Im\left\langle U({\bf r},\omega)\partial_{i}U^{*}({\bf r},\omega)\right\rangle. \label{energyflow}
\end{eqnarray}
As such, it is   {\it the averaged scattering force}, or {\it mean radiation pressure}, i.e.
\begin{eqnarray}
\left\langle \tilde{F}_{i}^{sc}\left(\mathbf{r},\omega\right)\right\rangle  & = & -2|{\bf e}(\omega)|^{2}\Im\alpha_{e}\Im\left\{ \partial_{i}^{(*)}W({\bf r},{\bf r},\omega)\right\} \nonumber \\
 & = & -2|{\bf e}(\omega)|^{2}\Im\alpha_{e}\Im\left\{ \left\langle U({\bf r},\omega)\partial_{i}U^{*}({\bf r},\omega)\right\rangle \right\} \nonumber \\
 & = & 2k|{\bf e}(\omega)|^{2}\Im\alpha_{e}\left\langle{\cal S}_{i}({\bf r},\omega)\right\rangle ,\label{Fsc}
\end{eqnarray}

 Eq. (\ref{Fsc}) manifests the correspondence of $\left\langle {\cal S} \right\rangle$ in this scalar formulation of the radiation pressure with the mean Poynting vector $\left\langle{\bf S}\right\rangle=(c/8\pi)\left\langle{\bf E}\times {\bf B}^{*}\right\rangle$   in the second term of Eq. (\ref{graddecomp}) acording to the vector representation. 

Notice that the {\it mean curl of electric spin density} which according to the third term of Eq. (\ref{graddecomp}) takes on the form
\begin{equation}
\left\langle \tilde{F}_{i}^{curl}\left(\mathbf{r},\omega\right)\right\rangle =2\Im\alpha_{e}\Im\left\{ e_{j}^{*}(\omega)e_{i}(\omega)\partial_{j}W(\mathbf{r},\mathbf{r},\omega)\right\} ,\label{Fcurl}
\end{equation}
 where $\partial_{i}$ means the derivative  in the  non-conjugated wavefunction, i.e., $\partial_{i}W(\mathbf{r},\mathbf{r},\omega)=\left\langle U^{*}({\bf r},\omega)\partial_{i}U({\bf r},\omega)\right\rangle $. Now, taking into account Maxwell's divergence equation $\nabla\cdot\mathbf{E}\left(\mathbf{r},\omega\right)=0$ in Eq. \eqref{Fcurl}, it is easy to demonstrate that $\left\langle \tilde{F}_{i}^{curl}(\mathbf{r},\omega)\right\rangle$
will be zero if $\Im\left\{ e_{j}^{*}(\omega)e_{i}(\omega)\right\} =0$; which evidently holds since ${\bf e}(\omega)$ is real. 

According to the Huygens-Fresnel principle \cite{MandelWolf,BornWolf,Perina}
the fluctuating field propagated from points ${\bf r}'$ of a surface
${\cal A}$ up to a point ${\bf r}$ is given by  
\begin{equation}
U({\bf r},\omega)=-\frac{ik}{2\pi}\int_{{\cal A}}U({\bf r}',\omega)\frac{e^{ikR}}{R}d^{2}r'.\label{UP}
\end{equation}
 where $R=|{\bf r}-{\bf r}'|$ and $k=2\pi/\lambda$, the wavelength
being $\lambda$ . Thus from Eq. (\ref{ftot1}) the
averaged force on a dipolar particle in ${\bf r}$ will be 
\begin{eqnarray}
\left\langle \tilde{F}_{i}\left(\mathbf{r},\omega\right)\right\rangle & = & -2\left(\frac{k}{2\pi}\right)^{2}|{\bf e}(\omega)|^{2}\Re\left\{ \alpha_{e}\int_{{\cal A}}\int_{{\cal A}}[ik+\frac{1}{R_{1}}]\frac{{\bf R_{1}}}{R_{1}}W({\bf r}_{1}',{\bf r}'_{2},\omega)\right.\nonumber \\
 & \times  & \left.\frac{e^{-ikR_{1}}}{R_{1}}\frac{e^{ikR_{2}}}{R_{2}}d^{2}r'_{1}d^{2}r'_{2}\right\} ,\label{FPQ}
\end{eqnarray}
 ${\bf R_{i}}={\bf r}-{\bf r'}_{i}$, $R_i=|{\bf r}-{\bf r}'_i|$, $(i=1,2)$. As mentioned
above, Eq. (\ref{FPQ}) exhibits the dependence of the mean force exerted by  the propagated field on its coherence properties on a surface ${\cal A}$.

 For instance, we consider the surface ${\cal A}$ being
composed of an opaque screen with two point holes, (see Fig. 1), so that  the random field wavefunction in ${\cal A}$
is: $U({\bf r}',\omega)=U({\bf q}_{1}, \omega)\delta({\bf r}'-{\bf q}_{1})+U({\bf q}_{2}, \omega)\delta({\bf r}'-{\bf q}_{2})$.
Then from Eqs. (\ref{UP}), (\ref{FPQ}), (\ref{Fgrad}) and (\ref{Fsc})
we obtain for the conservative and non-conservative force components on a particle at a point $P$ of position vector ${\bf r}$: 
\begin{eqnarray}
& & \left\langle \tilde{\mathbf{F}}^{grad}\left(\mathbf{r},\omega\right)\right\rangle \nonumber \\
 & = & -2\left(\frac{k}{2\pi}\right)^{2}|{\bf e}(\omega)|^{2}\Re\alpha_{e}\left\{ \left\langle |U({\bf q}_{1},\omega)|^{2}\right\rangle \frac{{\bf R}_{1}}{R_{1}^{4}}+\left\langle |U({\bf q}_{2},\omega)|^{2}\right\rangle \frac{{\bf R}_{2}}{R_{2}^{4}}\right.\nonumber \\
 & + & \frac{|W(\mathbf{q}_{1},\mathbf{q}_{2},\omega)|}{R_{1}R_{2}}\left[\left(\frac{{\bf R}_{1}}{R_{1}^{2}}+\frac{{\bf R}_{2}}{R_{2}^{2}}\right)\cos\left(k({\bf R}_{1}-{\bf R}_{2})+\alpha(\mathbf{q}_{1},\mathbf{q}_{2},\omega)\right)\right.\nonumber \\
 & + & \left.\left.\left(\frac{{\bf R}_{1}}{R_{1}}-\frac{{\bf R}_{2}}{R_{2}}\right)k\sin\left(k({\bf R}_{1}-{\bf R}_{2})+\alpha(\mathbf{q}_{1},\mathbf{q}_{2},\omega)\right)\right]\right\} ,\label{Younggrad}
\end{eqnarray}
 
\begin{eqnarray}
& & \left\langle \tilde{\mathbf{F}}^{sc}\left(\mathbf{r},\omega\right)\right\rangle \nonumber \\
 & = & 4\left(\frac{k}{2\pi}\right)^{2}|{\bf e}(\omega)|^{2}\Im\alpha_{e}\left\{ \left\langle |U({\bf q}_{1},\omega)|^{2}\right\rangle k\frac{{\bf R}_{1}}{R_{1}^{3}}+\left\langle |U({\bf q}_{2},\omega)|^{2}\right\rangle k\frac{{\bf R}_{2}}{R_{2}^{3}}\right.\nonumber \\
 & + & \frac{|W(\mathbf{q}_{1},\mathbf{q}_{2},\omega)|}{R_{1}R_{2}}\left[\frac{{\bf R}_{1}}{R_{1}^{2}}\left(kR_{1}\cos\left(k({\bf R}_{1}-{\bf R}_{2})+\alpha(\mathbf{q}_{1},\mathbf{q}_{2},\omega)\right)\right.\right.\nonumber \\
 & + & \left.\sin\left(k({\bf R}_{1}-{\bf R}_{2})+\alpha(\mathbf{q}_{1},\mathbf{q}_{2},\omega)\right)\right)\nonumber \\
 & + & \frac{{\bf R}_{2}}{R_{2}^{2}}\left(kR_{2}\cos\left(k({\bf R}_{1}-{\bf R}_{2})+\alpha(\mathbf{q}_{1},\mathbf{q}_{2},\omega)\right)\right.\nonumber \\
 & - & \left.\left.\left.\sin\left(k({\bf R}_{1}-{\bf R}_{2})+\alpha(\mathbf{q}_{1},\mathbf{q}_{2},\omega)\right)\right)\right]\right\}. \label{Youngsc}
\end{eqnarray}
 Denoting  ${\bf R}_{i}={\bf r}-{\bf q}_{i}$, ($i=1,2$),
and $\alpha\left(\mathbf{q}_{1},\mathbf{q}_{2},\omega\right)$ being the
phase of $W(\mathbf{q}_{1},\mathbf{q}_{2},\omega)$.

In the Fresnel and Fraunhofer  regions one may approximate $R_{1}\simeq R_{2}$ in the denominators of Eq. (\ref{Younggrad}). Also, for $kR_i \gg 1$, $(i=1,2)$,  the $cos$ terms are negligible versus the $sin$ terms and the gradient force has an interferencial $sin$ behavior, proportional to the difference: ${\bf R}_1-{\bf R}_2$.  On the other hand, the $sin$ terms of (\ref{Youngsc}) are negligible  versus the $cos$ terms, rendering a scattering force proportional to the intensity pattern.  This will be discussed again in Section 3 in connection with the configuration of Thompson and Wolf experiment, which replaces the two point holes of this schematic example by real apertures. In addition,  by dropping in Eqs. \eqref{Younggrad} and \eqref{Youngsc} the  corresponding factor constituted by  the real and imaginary part of the electric polarizability, we observe that the action on particles situated at  points $R_{i} \gg \lambda$ by the $\Im\alpha_e$-normalized repulsive scattering force  produced by each independent pinhole,  is much larger along ${\bf R}_i$ than that of the corresponding $\Re\alpha_e$-normalized attractive gradient force.

\section{Interference of two random waves: Degree of coherence and averaged optical
force}
 
In this section we address the force on a dipolar particle in the configuration of the classical two-aperture arrangement by Thompson and Wolf, employed in 1957 to observe and characterize the degree of coherence of a light wave \cite{thompson}.  As shown in the scheme of Fig. 1, the wavefield emitted by an incoherent source is brought by a lens $L_1$ to the mask $\cal A$ in $z=0$, containing two small circular apertures of radius $a$, centered at points $\mathbf{r}'_{1}=\left(\mathbf{q}_{1},0\right)$ and  $\mathbf{r}'_{2}=\left(\mathbf{q}_{2},0\right)$, respectively. A second lens $L_2$ sends the field diffracted in $\cal A$ to points $P$ of a screen ${\cal B}$ coinciding with its focal plane. The {\it degree of coherence} of the wave in ${\cal A}$: $\mu(\mathbf{q}_{1},\mathbf{q}_{2},\omega) $ $= W(\mathbf{q}_{1},\mathbf{q}_{2},\omega)/[W(\mathbf{q}_{1},\mathbf{q}_{1},\omega)W(\mathbf{q}_{2},\mathbf{q}_{2},\omega)]^{1/2} $ is expressed by means of the  Van-Cittert-Cernike theorem \cite{MandelWolf,BornWolf} in terms of the intensity exiting the incoherent source . In this way, we stablish the influence of the partial coherence in  ${\cal A}$ of the light emitted by the random incoherent source, on the optical force from the diffracted field upon a dipolar particle placed in ${\cal B}$.

In the vector theory of diffraction, within
the range of validity of the Kirchhoff approximation, the diffracted electric
vector produced by an aperture centered in ${\bf r}={\bf 0}$ in a screen $\cal A$   is expressed in the far zone as \cite{jackson1998classical,nietolibro}

\begin{equation}
\mathbf{\tilde{E}}(\mathbf{r},\omega)=\frac{ie^{ikr}}{2\pi r}\mathbf{k}\times\int_{\cal A}\mathbf{n}\times\mathbf{\tilde{E}}^{(i)}(\mathbf{r'},\omega)e^{-i\mathbf{k}\cdot\mathbf{r}'}ds',\label{jackson}
\end{equation}
 where $\mathbf{k}=k\mathbf{s}=(2\text{\ensuremath{\pi}}/\lambda)\mathbf{s}$,
$\mathbf{s}=\mathbf{r}/r$ is a unit vector in the direction of observation
$\mathbf{r}=\left(x,y,z\right)$, \textbf{$\mathbf{r}'$ } denotes a coordinate in the aperture whose element of surface area is $ds'$, and $\mathbf{n}$ is the unit outward
 normal to $ds'$. The time-dependence $e^{-i\omega t}$ is understood,
and  $\mathbf{\tilde{E}}^{(i)}(\mathbf{r'})={\bf e}^{(i)}(\omega)\exp(ik{\bf n}^{(i)}\cdot{\bf r}')$,
($|{\bf n}^{(i)}|=1$, $\Im {\bf e}^{(i)}=0$),  is the electric field incident on the mask $\mathcal{A}$. 

\begin{figure}
\begin{centering}
\includegraphics[scale=0.5]{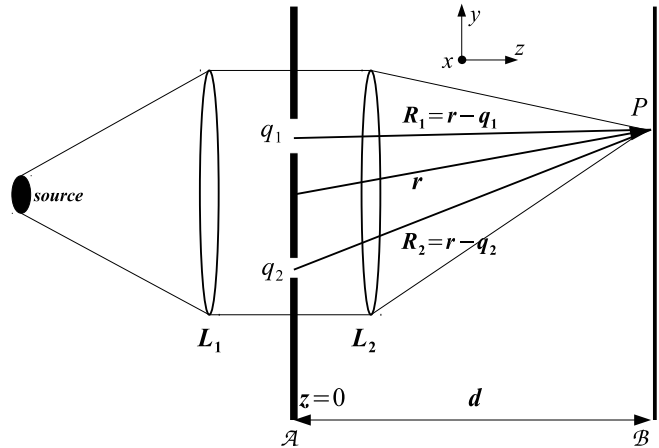} 
\par\end{centering}

\caption{Schematics of the configuration for  observing interference  at points $P=\left(x,y,d\right)$ of a screen  ${\cal B}$ by diffraction of light, propagated from an incoherent source, at two apertures in ${\cal A}$, centered at points
$\mathbf{r}'_{1}=\left(\mathbf{q}_{1},0\right)$ and  $\mathbf{r}'_{2}=\left(\mathbf{q}_{2},0\right)$, respectively.
$\mathbf{q}_{1}=\left(x_{1},y_{1}\right)$, $\mathbf{q}_{2}=\left(x_{2},y_{2}\right)$. $2h=|\mathbf{q}_{1}-\mathbf{q}_{2}|$.\label{Fig1}}
\end{figure}

 Assuming ${\bf n}^{(i)}=(0,0,1)$, a flat opaque screen in $z=0$  with a circular aperture of  center ${\bf r}'={\bf 0}$ and radius $a$  produces according to Eq. (\ref{jackson}) the diffracted field \cite{BornWolf, goodmam1996} 
\begin{equation}
\mathbf{\tilde{E}}(\mathbf{r},\omega)=U\left(\mathbf{r},\omega\right)\mathbf{e}(\omega),\label{E},
\end{equation}
 where
\begin{equation}
U(\mathbf{r},\omega)=e^{ikz}e^{i\frac{k}{2z}(x^{2}+y^{2})}\left(\frac{\pi a^{2}}{i\lambda z}\right)\left(2\frac{J_{1}(v_0)}{v_0}\right).\label{V}
\end{equation}
 In Eqs. (\ref{E}) and (\ref{V}) we have written $\mathbf{e}(\omega)=\mathbf{s}\times\left(\mathbf{n}\times\mathbf{e}^{\left(i\right)}(\omega)\right)$  and $v_{0}=ka\sqrt{x^{2}+y^{2}}/z$, respectively.

Next we study the effect of the degree of coherence in  ${\cal A}$ of the wavefield propagated from the chaotic source on the mean force upon a dipolar particle in the Fraunhofer zone. Hence we shall address the consequences of
the correlation between the field at the two circular apertures in $z=0$, centered at $\mathbf{q}_{1}\equiv(0,h,0)$
and $\mathbf{q}_{2}\equiv(0,-h,0)$, as shown in Fig. 1. To
this end, we evaluate the field  at an arbitrary point $P$ in the plane
\textbf{$\mathcal{B}$ } at $z=d$ where the particle is situated, produced on diffraction in ${\cal A}$, (cf. Fig. 1) within the Kirchhoff approximation:
\begin{equation}
U(\mathbf{r},\omega)=e^{ikd}e^{i\frac{k}{2d}(x^{2}+y^{2})}\left(\frac{\pi a^{2}}{i\lambda d}\right)\left(2\frac{J_{1}\left(v_{0}\right)}{v_{0}}\right)\left[U(\mathbf{q}_{1},\omega)e^{-i\frac{khy}{d}}+U(\mathbf{q}_{2},\omega)e^{i\frac{khy}{d}}\right],\label{V_dos_agujeros}
\end{equation}
 where $U(\mathbf{q}_{i},\omega)$ \textbf{$\left(i=1,2\right)$}
is the complex amplitude of the random wavefield  at $\mathbf{q}_{1}$ and $\mathbf{q}_{2}$ emitted by the fluctuating source. If as in Thompson and Wolf experiment \cite{thompson}, one has that $\left\langle \left|U(\mathbf{q}_{1},\omega)\right|^{2}\right\rangle \thickapprox\left\langle \left|U(\mathbf{q}_{2},\omega)\right|^{2}\right\rangle =I_{0}$,
the observed mean intensity  at $P$ then is known to be
\begin{eqnarray}
\left\langle I\left(\mathbf{r},\omega\right)\right\rangle  & = & \left\langle \mathbf{\tilde{E}}(\mathbf{r},\omega) \cdot \mathbf{\tilde{E}}^{*}(\mathbf{r},\omega)\right\rangle \nonumber \\
 & = & 2I_{0}\left(\frac{\pi a^{2}\left|\mathbf{e}\left(\omega\right)\right|}{\lambda d}\right)^{2}\left(2\frac{J_{1}\left(v_{0}\right)}{v_{0}}\right)^{2}\nonumber \\
 & \times & \left[1+\left|\mu(\mathbf{q}_{1},\mathbf{q}_{2},\omega)\right|{\displaystyle \textrm{cos}}\left(\phi(\mathbf{q}_{1},\mathbf{q}_{2},\omega)+\frac{2khy}{d}\right)\right]\label{Int_reducida}
\end{eqnarray}

The factor $2khy/d$ represents the path difference $\left|\mathbf{R}_{1}-\mathbf{R}_{2}\right|$,
(see Fig. 1). $\phi(\mathbf{q}_{1},\mathbf{q}_{2},\omega)$
is the phase of the degree of coherence: $\mu(\mathbf{q}_{1},\mathbf{q}_{2},\omega)=2J_{1}(u)/u$ in $z=0$. Where $u=2\pi \rho h/(\lambda \Delta)$. $\rho$ being the radius of the source, assumed planar and circular, and $\Delta$ denoting the distance between the source and $L_1$  \cite{thompson}.
 
 The interference law of Eq. \eqref{Int_reducida} is well known \cite{thompson,BornWolf}. We shall perform calculations of the force with the same parameters as in Ref. \cite{thompson}, namely: $\lambda=579\, nm$, $2h=6\, mm$, $a=0.7 mm$, $d=1.5\, m$. We  consider $\phi(\mathbf{q}_{1},\mathbf{q}_{2},\omega)=0$
and  we will initially normalize the results to the polarizability, so that we will make $\Re\alpha_{e}=\Im\alpha_{e}=1$; this allows us to obtain an estimation of
the relative strengths of the different force components due to diffraction, independently of the polarizability.

In this far zone, the {\it gradient force} is governed by the expression (\ref{Fgrad}) applied to the mean  intensity (\ref{Int_reducida}). Since the apertures are aligned along the $y-$axis, the $y$ component for $kR_{i}\rightarrow\infty$ $\left(i=1,2\right)$ is obtained after a long but straightforward algebra 
\begin{eqnarray}
\left\langle \tilde{F}_{y}^{grad}\right\rangle  & \approx & -4\Re\alpha_{e}I_{0}\left(\frac{\pi a^{2}\left|\mathbf{e}(\omega)\right|}{\lambda d}\right)^{2}\left(2\frac{J_{1}\left(v_{0}\right)}{v_{0}}\right)^{2}\frac{hk}{d}\nonumber \\
 & \times & \left|\mu(\mathbf{q}_{1},\mathbf{q}_{2},\omega)\right|{\displaystyle \textrm{sin}}\left(\phi(\mathbf{q}_{1},\mathbf{q}_{2},\omega)+\frac{2khy}{d}\right),\label{FgradY}
\end{eqnarray}
which agrees with the remark at the end of Section 2 concerning Eq. (\ref{Younggrad}). 

\begin{figure}
\begin{centering}
\includegraphics{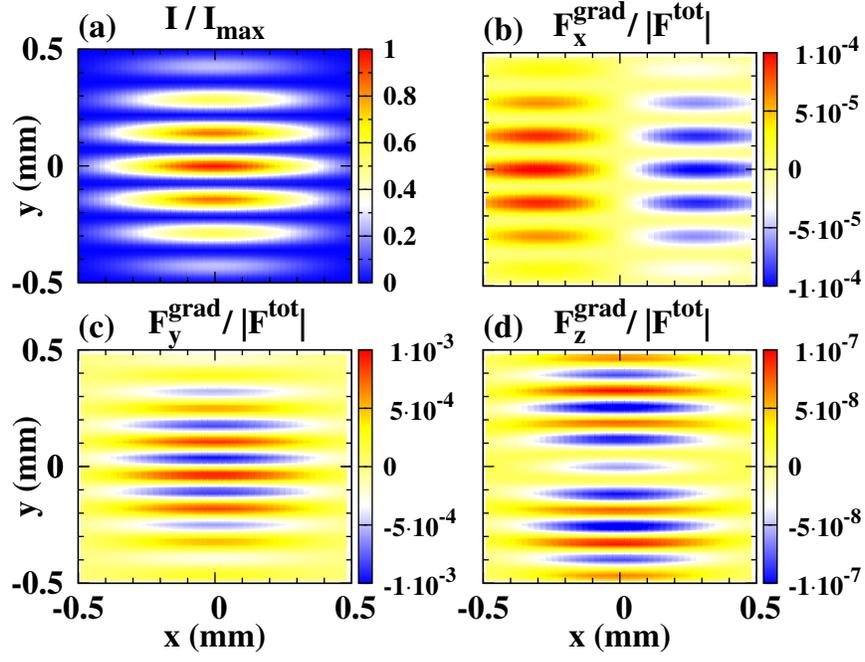} 
\par\end{centering}

\caption{Spatial distributions  in the $XY$ plane of the intensity and the normalized gradient force components for $\left|\mu(\mathbf{q}_{1},\mathbf{q}_{2},\omega)\right|=1$.
(a) Normalized mean intensity $\left\langle I\right\rangle $. (b) $\left\langle \tilde{F}_{x}^{grad}\right\rangle $. (c) $\left\langle \tilde{F}_{y}^{grad}\right\rangle $. (d) $\left\langle \tilde{F}_{z}^{grad}\right\rangle $.
All values are calculated on a dipolar particle at the screen plane $\mathcal{B}$, placed
at distance $z=d=1.5\, m$ from the aperture screen $\mathcal{A}$. The force components are normalized to  $\Re\alpha_e$ and to the magnitude of the total mean force $\left|\left\langle \mathbf{\tilde{F}}^{tot}\right\rangle \right|=\left|\left\langle \mathbf{\tilde{F}}^{grad}\right\rangle +\left\langle \mathbf{\tilde{F}}^{sc}\right\rangle \right|$.}
\end{figure}

This expression is just the derivative of Eq. \eqref{Int_reducida}
with respect to $y$, assuming that the factor outside the brackets in (\ref{Int_reducida}) is constant, (although this is not strictly true, the terms yielded by the y-derivative of this factor become negligible, as shown in the Appendix A). The other two components: $\left\langle \tilde{F}_{x}^{grad}\right\rangle $ and $\left\langle \tilde{F}_{z}^{grad}\right\rangle $ are similarly obtained in the Appendix A.

Fig. 2 shows the interference pattern of $\left\langle I \right\rangle$,  normalized to its maximum, at the screen ${\cal B}$  for $\left|\mu(\mathbf{q}_{1},\mathbf{q}_{2},\omega)\right|=1$, as well as the spatial distribution of the three  components of the mean gradient force on a dipolar particle in ${\cal B}$ due to this distribution of light. To see the relative weight of each Cartesian component, we normalize it to  the magnitude of the total mean force $\left|\left\langle \mathbf{\tilde{F}}^{tot}\right\rangle \right|=\left|\left\langle \mathbf{\tilde{F}}^{grad}\right\rangle +\left\langle \mathbf{\tilde{F}}^{sc}\right\rangle \right|$. We also observe an interference pattern along OY  in each  component of this conservative force,  $\left\langle \tilde{F}_{z}^{grad}\right\rangle $ being much smaller than the other two. In addition, Fig. 2(b)  exhibits an oscillatory modulation of $\left\langle \tilde{F}_{x}^{grad}\right\rangle $
along $OX$, (cf.  Appendix A).

\begin{figure}
\begin{centering}
\includegraphics[scale=0.9]{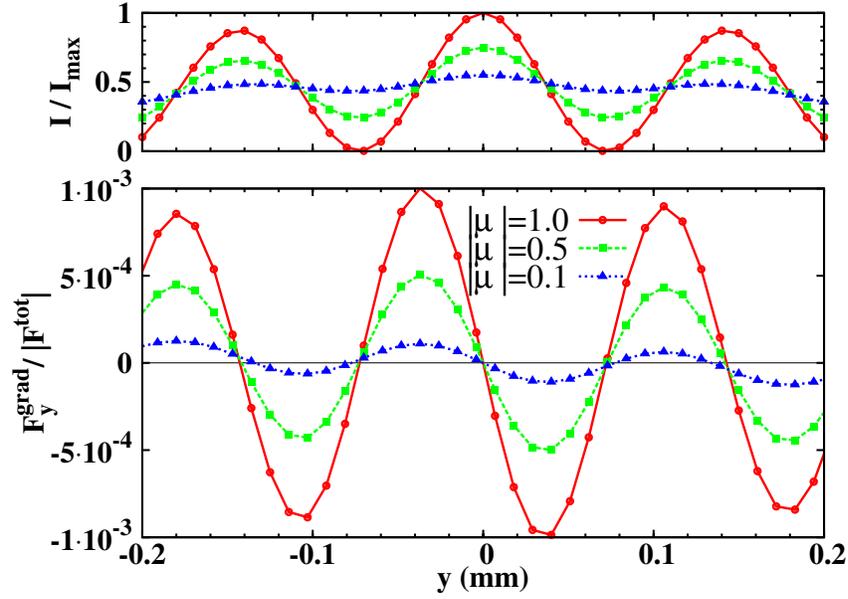} 
\par\end{centering}

\caption{(a) Normalized mean intensity $\left\langle I \right\rangle$. (b) Normalized gradient force component $\left\langle \tilde{F}_{y}^{grad}\right\rangle $
for different values of $\left|\mu(\mathbf{q}_{1},\mathbf{q}_{2},\omega)\right|$. }
\end{figure}

We remark that in the limiting case $\left|\mu(\mathbf{q}_{1},\mathbf{q}_{2},\omega)\right|=0$, $\left\langle \tilde{F}_{y}^{grad}\right\rangle $ is just proportional to $\partial_{y}\left(2J_{1}\left(v_{0}\right)/v_{0}\right)^{2}$
and the interference effect disappears, as it should. Since this $y$ derivative was neglected versus the term kept in Eq. (\ref{FgradY}), (see also Appendix A), the values of  $\left\langle \tilde{F}_{y}^{grad}\right\rangle $ then are practically zero compared to those due to a partially coherent wave. This is seen in Fig.3.
The intensity pattern, which acts as a potential distribution for the illuminated particle, is shifted by $\pi/2$ with respect to that of the conservative force $\left\langle \tilde{F}_{y}^{grad}\right\rangle $, whose oscillation amplitude progressively  diminishes to zero as the value of $\left|\mu(\mathbf{q}_{1},\mathbf{q}_{2},\omega)\right|$
decreases.  This behavior of the conservative  force constitutes the basic mechanism of an optical tweezer with several equilibrium positions of the particle along the lines in the screen ${\cal B}$ where $\left\langle I\right\rangle $ is maximum, .  Such points occur along $OY$ at $x=0$ , [cf. Figs. 2(a) and 2(c)], and are precisely those where $\left\langle \tilde{F}_{x}^{grad}\right\rangle =0$, [cf. Fig. 2(b)].

We next address the {\it scattering force} on a small particle in an arbitrary point of the screen $\mathcal{B}$, obtained on introducing
Eq. \eqref{V_dos_agujeros} into Eq. \eqref{Fsc}, (see Appendix B). The  scattering and gradient force $x-$components are of similar magnitude, but of signs opposite to each other; this is seen on comparing Fig. 4 (a) with Fig. 2 (b). By contrast, the $y$-component of $\left\langle \tilde{F}^{sc}\right\rangle$ is one order of magnitude smaller than its homologous of the gradient force, suffering a sharp change of sign at $y=0$. However,  $\left\langle \tilde{F}_{z}^{sc}\right\rangle $ which is given by (see the Appendix B) 
\begin{eqnarray}
\left\langle \tilde{F}_{z}^{sc}\right\rangle  & \approx & 4k\Im\alpha_{e}I_{0}\left(\frac{\pi a^{2}\left|\mathbf{e}(\omega)\right|}{\lambda d}\right)^{2}\left(2\frac{J_{1}\left(v_{0}\right)}{v_{0}}\right)^{2}\nonumber \\
 & \times & \left[1+\left|\mu(\mathbf{q}_{1},\mathbf{q}_{2},\omega)\right|{\displaystyle \textrm{cos}}\left(\phi(\mathbf{q}_{1},\mathbf{q}_{2},\omega)+\frac{2khy}{d}\right)\right]\nonumber \\
 & \approx & 2k\Im\alpha_{e} \left\langle I(\mathbf{r},\omega) \right\rangle, \label{Fsz}
\end{eqnarray}
 is seven orders of magnitude larger than the corresponding conservative force  $\left\langle \tilde{F}_{z}^{grad}\right\rangle $, [compare Fig. 4 (c) with Fig. 2 (d)]. This, which is in accordance with the remark of the last paragraph of Section 2 concerning Eqs. \eqref{Younggrad}
and \eqref{Youngsc} for waves from two pinholes, stems from the proportionality of $\left\langle F_{z}^{sc}\right\rangle $ to the Poynting vector \cite{MNVOpex} and hence to the mean scattered intensity in the far zone; [observe that the normalized $\left\langle F_{z}^{sc}\right\rangle $ of Fig. 4(c) is identical to the normalized mean intensity  $\left\langle I\right\rangle $ of Fig. 2 (a)]. As a consequence of the conservation of momentum, the particle is pushed towards $z>d$ (see Fig. 1).  {\it The ratio between the maximum values of the gradient and scattering force components}, (see Eqs. (\ref{FgradY}), (\ref{Fsz}) and the Appendix A and B), $\left\langle \tilde{F}_{z}^{grad}\right\rangle /\left\langle \tilde{F}_{z}^{sc}\right\rangle =-(y/d)\left\langle \tilde{F}_{y}^{grad}\right\rangle /\left\langle \tilde{F}_{z}^{sc}\right\rangle =\left(yh/d^2\right)\left(\left|\mu\right|/1+\left|\mu\right|\right)$ explains the difference between the magnitudes of these force components.

\begin{figure}
\begin{centering}
\includegraphics{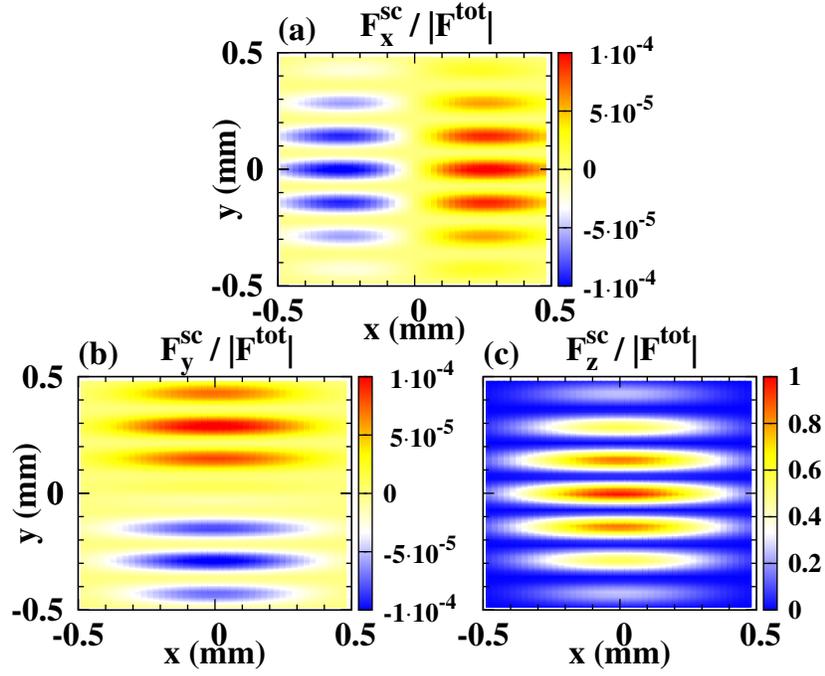} 
\par\end{centering}

\caption{Spatial distribution in the $XY$  plane of the normalized scattering force components  for $\left|\mu(\mathbf{q}_{1},\mathbf{q}_{2},\omega)\right|=1$.  The normalization factor is $\Im\alpha_{e} \left|\left\langle \tilde{\mathbf{F}}^{tot}\right\rangle \right|$.  (a) $\left\langle F_{x}^{sc}\right\rangle $.
(b) $\left\langle F_{y}^{sc}\right\rangle $. (c) $\left\langle F_{z}^{sc}\right\rangle $. All values are calculated on a dipolar particle in the screen plane $\mathcal{B}$, placed at distance $z=d=1.5\, m$ from the aperture screen $\mathcal{A}$. }

\label{Fig4} 
\end{figure}
The much larger strength of the normalized scattering force may prevent the lateral manipulation of the particle in ${\cal B}$.  If this were the case, it may be overcame with a scheme analogous to that employed in holographic optical tweezers \cite{curtis2002dynamic}.

Summing up, we observe  that the Young experiment configuration shows us fundamental characteristics of the optical force components, which in this system allow a scalar formulation and thus  yield  no {\it curl} component. {\it The mean  scattering force is proportional to the  mean scattered intensity, its longitudinal $z$-component being several orders of magnitude larger than its $x$ and $y$ components}. On the other hand, {\it  the $y$ and $z$-components of the gradient force are  proportional to the magnitude of the degree of coherence, thus becoming zero for incoherent light, and hence their value oscillates and decreases  as $J_{1}(2\pi\rho h/\lambda\Delta)/\left(2\pi\rho h/\lambda\Delta\right)$ versus the distance $h$ between apertures} \cite{thompson}.

\subsection{Effect of the electric polarizability on the mean optical force}
We have so far estimated  the different components of
the mean force by only considering the configuration of the diffracted waves; namely, by normalizing them  to the particle polarizability.
However, it is worth remarking that since their actual strengths,  observed in an experiment,  are proportional to $\Re\alpha_{e}$ (gradient force)
ant to $\Im\alpha_{e}$ (scattering force), the relative values of
these two parts of $\alpha_{e}$ should greatly influence the magnitude of these
forces. Notice that although we have concluded that $\left\langle \tilde{F}_{y}^{grad}\right\rangle \ll\left\langle \tilde{F}_{z}^{sc}\right\rangle $
when they are normalized to $\Re\alpha_{e}$ and $\Im\alpha_{e}$,
respectively, in most cases pertaining to dielectric particles one
has that $\Re\alpha_{e}\gg\Im\alpha_{e}$, except in the presence
of Mie electric and/or magnetic  \cite{MNVOpex,MNVJOSA010} or plasmon \cite{Arias03} resonances.

For a small spherical particle of radius $r_{0}$, with relative 
permittivity $\varepsilon_{p}$, in the Rayleigh limit $(kr_{0}\ll1)$,
we adopt the expression for the dynamic electric polarizability \cite{MNVOpex,MNVJOSA010}:
\begin{equation}
\alpha_{e}=\alpha_{e}^{\left(0\right)}\left(1-i\frac{2}{3}k^{3}\alpha_{e}^{\left(0\right)}\right)^{-1},\label{polarizabilidad_electrica}
\end{equation}
 $\alpha_{e}^{\left(0\right)}$ being the static polarizabilty 
\begin{equation}
\alpha_{e}^{\left(0\right)}=r_{0}^{3}\frac{\varepsilon_{p}-1}{\varepsilon_{p}+2}\label{polarizabilidad_estatica}
\end{equation}

\begin{figure}
\begin{centering}
\includegraphics{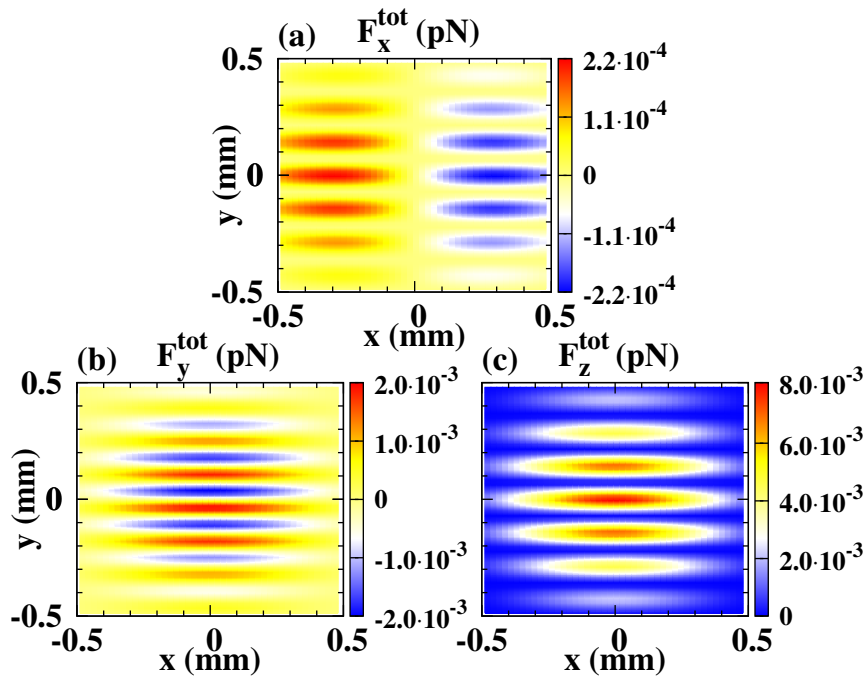} 
\par\end{centering}

\caption{Spatial distribution, in pN,  of the total  force Cartesian components on a dielectric particle with $r_{0}=25\, nm$
and $\varepsilon_{p}=2.25$, in the screen plane $\mathcal{B}$, placed at distance $z=d=1.5\, m$ from the aperture mask $\mathcal{A}$, $\left|\mu(\mathbf{q}_{1},\mathbf{q}_{2},\omega)\right|=1$.  No normalization is done.  (a) $\left\langle F_{x}^{tot}\right\rangle $. (b) $\left\langle F_{y}^{tot}\right\rangle $. (c) $\left\langle F_{z}^{tot}\right\rangle $.  }

\label{Fig5} 
\end{figure}

As an illustration, we consider a  dielectric particle with $r_{0}=25\, nm$
and $\varepsilon_{p}=2.25$. With these data, we observe as 
mentioned above that $\Re\alpha_{e}=4593\, nm^{3}\gg\Im\alpha_{e}=17\, nm^{3}$. The illumination that reaches each aperture is assumed with a magnitude
of the Poynting vector $(c/2)I_{0}\left|\mathbf{e}\left(\omega\right)\right|^{2}=10^{12}\, W/m^{2}$.
Fig. 5 shows the corresponding different components of the total force, (this time of course  without performing any normalization).

As  seen, the patterns of Fig. 5 (a) and Fig. 5 (b) are equal to those of
Fig. 2 (b) and Fig. 2 (c) respectively; this implying that the scattering force is negligible compared to the gradient force along $OX$ and $OY$. However,  although
$\Re\alpha_{e} \gg \Im\alpha_{e}$, this is not
enough for $\left\langle \tilde{F}_{z}^{grad}\right\rangle $ to exceed
$\left\langle \tilde{F}_{z}^{sc}\right\rangle $, (remember that we obtained a
difference of seven orders of magnitude between these two normalized $z$-components), therefore the contribution of  $\left\langle \tilde{F}_{z}^{grad}\right\rangle $ to  $\left\langle \tilde{F}_{z}^{tot}\right\rangle$  is  negligible by four orders of magnitude.  Notwithstanding, it is important for trapping purposes that the $y$ component of the force   $\left\langle \tilde{F}_{y}^{tot}\right\rangle$, which is of conservative nature,    is of the same order of magnitude as the non-conservative $z$-force $\left\langle \tilde{F}_{z}^{tot}\right\rangle$. As the coherence diminishes, Fig. 3(b) gives an assessment of the corresponding decrease to be expected in both $\left\langle \tilde{F}_{x}^{tot}\right\rangle$ and $\left\langle \tilde{F}_{y}^{tot}\right\rangle$ from their values in  Figs.  5(a) and 5(b).

\section{Conclusion}
We have established a theory of the averaged optical force exerted by a partially coherent stationary and ergodic  electromagnetic field on a dipolar particle. Although we have put forward expressions for the force acting on magnetodielectric particles, we have illustrated the  effect of partial coherence  on a particle that has electric polarizability only. This has been carried out by calculating the influence that either the cross-spectral density and the degree of coherence  of the fluctuating field at a certain plane of propagation,  have on the force components due to light propagated  beyond that plane.

To this end, we have made use of the classical two-apertures interference configuration of Young to calculate the force on a dipolar particle in the Fraunhofer zone. Hence, in correspondence with  the classical study on coherence by Thompson and Wolf \cite{thompson}, we have calculated the  force components  and their dependence on the degree of coherence of the fluctuating field at the plane of the  apertures. The result is quite interesting; it shows a fringe pattern spatial distribution in each  Cartesian component of the conservative and non-conservative forces, and illustrates the way in which they compete with each other to the resulting total forces.

The case of  magnetodielectric particles may be similarly studied. Then, apart from adding to the above the pure magnetic force whose strength  depends on the magnetic polarizability of the particle, an electric-magnetic interference force component, of opposite sign to the former, has to be summed to the latter. The relative weight of these latter two forces, as well as their dependence on  the degree of coherence of the  fluctuating field, should be evaluated. Of special interest will be the case of nanometric size high index, or semiconductor, spheres, illuminated in the near-infrared at which both their electric and magnetic dipole Mie resonances are excited \cite{GarciaEtxarri11Anistropic, MNVJOSA010}.

Likewise, although we have emphasized our illustration of Section 3 with an scalar theory, it will be of interest to calculate these optical forces in terms of the degree of polarization of the fluctuating fields by making use of the full electromagnetic model of Section 2.

We believe that this study will motivate further research on particle manipulation by fluctuating wavefields, both in the context of partially coherent waves and in the more general of random wavefields \cite{radiophysics, Ripoll01Photon, Riley00Three, garcia97Transition} , which should permit to work with light and other electromagnetic wave sources like partially fluctuating nanoantennas of limited coherence or correlation length or  thermal sources in which partial coherence is undeced. This will be of special importance at the nanometric scale in the near field, where new consequences are to be found.

%\bibliographystyle{osajnl}  %%%%the style of optics express
%\bibliography{Time-averagedforce}

\newpage
\section*{Appendix}
\appendix
\numberwithin{equation}{section}

\section{Analytical expressions of the gradient force}

\textit{Calculation of $\tilde{F}_{x}^{grad}$}

\begin{eqnarray}\label{analFgx}
&&\left\langle \tilde{F}_{x}^{grad}\right\rangle \nonumber\\
&=&-4\Re\alpha_{e}I_{0}\left(\frac{ak}{zv_{0}}\right)^{2}\left(\frac{\pi a^{2}\left|\mathbf{e}\left(\omega\right)\right|^{2}}{\lambda z}\right)^{2}x\left[\left(2\frac{J_{1}\left(v_{0}\right)}{v_{0}}\right)^{2}+2\frac{J_{1}\left(v_{0}\right)}{v_{0}}\right.\nonumber\\
&\times &\left.\left[J_{0}\left(v_{0}\right)-J_{2}\left(v_{0}\right)\right]\right]\left[1+\left|\mu\left(\mathbf{q}_{1},\mathbf{q}_{2},\omega\right)\right|\text{cos}\left(\phi\left(\mathbf{q}_{1},\mathbf{q}_{2},\omega\right)+\frac{2khy}{z}\right)\right].\nonumber\\ 
&&
\end{eqnarray}
$v_{0}=ka\sqrt{x^{2}+y^{2}}/z$. Observe that in this equation, none of the two terms may be neglected.\\\\
\textit{Calculation of $\tilde{F}_{y}^{grad}$}

\begin{eqnarray}\label{analFgy}
&&\left\langle \tilde{F}_{y}^{grad}\right\rangle \nonumber\\
&=&4\Re\alpha_{e}I_{0}\left(\frac{ak}{zv_{0}}\right)^{2}\left(\frac{\pi a^{2}\left|\mathbf{e}\left(\omega\right)\right|^{2}}{\lambda z}\right)^{2}y\left[\left(2\frac{J_{1}\left(v_{0}\right)}{v_{0}}\right)^{2}+2\frac{J_{1}\left(v_{0}\right)}{v_{0}}\right.\nonumber\\
&\times&\left.\left[J_{0}\left(v_{0}\right)-J_{2}\left(v_{0}\right)\right]\right]\left[1+\left|\mu\left(\mathbf{q}_{1},\mathbf{q}_{2},\omega\right)\right|\text{cos}\left(\phi\left(\mathbf{q}_{1},\mathbf{q}_{2},\omega\right)+\frac{2khy}{z}\right)\right]\nonumber\\
&-&4\Re\alpha_{e}I_{0}\left(\frac{\pi a^{2}\left|\mathbf{e}\left(\omega\right)\right|^{2}}{\lambda z}\right)^{2}\frac{hk}{z}\left(2\frac{J_{1}\left(v_{0}\right)}{v_{0}}\right)^{2}\nonumber\\
&\times&\left|\mu\left(\mathbf{q}_{1},\mathbf{q}_{2},\omega\right)\right|\text{sin}\left(\phi\left(\mathbf{q}_{1},\mathbf{q}_{2},\omega\right)+\frac{2khy}{z}\right)\nonumber\\
&=&\left\langle \tilde{F}_{x}^{grad}\right\rangle \frac{y}{x}-4\Re\alpha_{e}I_{0}\left(\frac{\pi a^{2}\left|\mathbf{e}\left(\omega\right)\right|^{2}}{\lambda z}\right)^{2}\frac{hk}{z}\left(2\frac{J_{1}\left(v_{0}\right)}{v_{0}}\right)^{2}\nonumber\\
&\times&\left|\mu\left(\mathbf{q}_{1},\mathbf{q}_{2},\omega\right)\right|\text{sin}\left(\phi\left(\mathbf{q}_{1},\mathbf{q}_{2},\omega\right)+\frac{2khy}{z}\right).
\end{eqnarray}
Hence
\begin{eqnarray}\label{analFgy2}
\left\langle \tilde{F}_{y}^{grad}\right\rangle &\approx&-4\Re\alpha_{e}I_{0}\left(\frac{\pi a^{2}\left|\mathbf{e}\left(\omega\right)\right|^{2}}{\lambda z}\right)^{2}\frac{hk}{z}\left(2\frac{J_{1}\left(v_{0}\right)}{v_{0}}\right)^{2}\nonumber\\
&\times &\left|\mu\left(\mathbf{q}_{1},\mathbf{q}_{2},\omega\right)\right|\text{sin}\left(\phi\left(\mathbf{q}_{1},\mathbf{q}_{2},\omega\right)+\frac{2khy}{z}\right).
\end{eqnarray}\\\\
\textit{Calculation of $\tilde{F}_{z}^{grad}$}

\begin{eqnarray}\label{analFgz}
&&\left\langle \tilde{F}_{z}^{grad}\right\rangle \nonumber\\
&=&-4\Re\alpha_{e}I_{0}\left(\frac{\pi a^{2}\left|\mathbf{e}\left(\omega\right)\right|^{2}}{\lambda z}\right)^{2}\frac{1}{z}\left(2\frac{J_{1}\left(v_{0}\right)}{v_{0}}\right)^{2}\left[J_{0}\left(v_{0}\right)-J_{2}\left(v_{0}\right)\right]\nonumber\\
&\times&\left[1+\left|\mu\left(\mathbf{q}_{1},\mathbf{q}_{2},\omega\right)\right|\text{cos}\left(\phi\left(\mathbf{q}_{1},\mathbf{q}_{2},\omega\right)+\frac{2khy}{z}\right)\right]\nonumber\\
&+&4\Re\alpha_{e}I_{0}\left(\frac{\pi a^{2}\left|\mathbf{e}\left(\omega\right)\right|^{2}}{\lambda z}\right)^{2}\frac{hky}{z^{2}}\left(2\frac{J_{1}\left(v_{0}\right)}{v_{0}}\right)^{2}\nonumber\\
&\times&\left|\mu\left(\mathbf{q}_{1},\mathbf{q}_{2},\omega\right)\right|\text{sin}\left(\phi\left(\mathbf{q}_{1},\mathbf{q}_{2},\omega\right)+\frac{2khy}{z}\right).
\end{eqnarray}
Therefore
\begin{eqnarray}\label{analFgz2}
\left\langle \tilde{F}_{z}^{grad}\right\rangle &\approx&4\Re\alpha_{e}I_{0}\left(\frac{\pi a^{2}\left|\mathbf{e}\left(\omega\right)\right|^{2}}{\lambda z}\right)^{2}\frac{hky}{z^{2}}\left(2\frac{J_{1}\left(v_{0}\right)}{v_{0}}\right)^{2}\nonumber\\
&\times&\left|\mu\left(\mathbf{q}_{1},\mathbf{q}_{2},\omega\right)\right|\text{sin}\left(\phi\left(\mathbf{q}_{1},\mathbf{q}_{2},\omega\right)+\frac{2khy}{z}\right)\nonumber\\
&=&-\frac{y}{z}\left\langle \tilde{F}_{y}^{grad}\right\rangle .
\end{eqnarray}

\section{Analytical expressions of the scattering force}

\textit{Calculation of $\tilde{F}_{x}^{sc}$}
\begin{eqnarray}\label{analFsx}
\left\langle \tilde{F}_{x}^{sc}\right\rangle &=&4\Im\alpha_{e}I_{0}\left(\frac{\pi a^{2}\left|\mathbf{e}\left(\omega\right)\right|^{2}}{\lambda z}\right)^{2}\frac{kx}{z}\left(2\frac{J_{1}\left(v_{0}\right)}{v_{0}}\right)^{2}\nonumber\\
&\times&\left[1+\left|\mu\left(\mathbf{q}_{1},\mathbf{q}_{2},\omega\right)\right|\text{cos}\left(\phi\left(\mathbf{q}_{1},\mathbf{q}_{2},\omega\right)+\frac{2khy}{z}\right)\right].
\end{eqnarray}\\\\
\textit{Calculation of $\tilde{F}_{y}^{sc}$}

\begin{eqnarray}\label{analFsy}
\left\langle \tilde{F}_{x}^{sc}\right\rangle &=&4\Im\alpha_{e}I_{0}\left(\frac{\pi a^{2}\left|\mathbf{e}\left(\omega\right)\right|^{2}}{\lambda z}\right)^{2}\frac{ky}{z}\left(2\frac{J_{1}\left(v_{0}\right)}{v_{0}}\right)^{2}\nonumber\\
&\times&\left[1+\left|\mu\left(\mathbf{q}_{1},\mathbf{q}_{2},\omega\right)\right|\text{cos}\left(\phi\left(\mathbf{q}_{1},\mathbf{q}_{2},\omega\right)+\frac{2khy}{z}\right)\right].
\end{eqnarray}
\textit{Calculation of $\tilde{F}_{z}^{sc}$}

\begin{eqnarray}\label{analFsz}
\left\langle \tilde{F}_{z}^{sc}\right\rangle &=&4k\Im\alpha_{e}I_{0}\left(\frac{\pi a^{2}\left|\mathbf{e}\left(\omega\right)\right|^{2}}{\lambda z}\right)^{2}\left(2z^{2}-x^{2}-y^{2}\right)\left(2\frac{J_{1}\left(v_{0}\right)}{v_{0}}\right)^{2}\nonumber\\
&\times&\left[1+\left|\mu\left(\mathbf{q}_{1},\mathbf{q}_{2},\omega\right)\right|\text{cos}\left(\phi\left(\mathbf{q}_{1},\mathbf{q}_{2},\omega\right)+\frac{2khy}{z}\right)\right]\nonumber\\
&\simeq&4k\Im\alpha_{e}I_{0}\left(\frac{\pi a^{2}\left|\mathbf{e}\left(\omega\right)\right|^{2}}{\lambda z}\right)^{2}\left(2\frac{J_{1}\left(v_{0}\right)}{v_{0}}\right)^{2}\nonumber\\
&\times&\left[1+\left|\mu\left(\mathbf{q}_{1},\mathbf{q}_{2},\omega\right)\right|\text{cos}\left(\phi\left(\mathbf{q}_{1},\mathbf{q}_{2},\omega\right)+\frac{2khy}{z}\right)\right]\nonumber\\
&=&2k\Im\alpha_{e}\left\langle I\left(\mathbf{r},\omega\right)\right\rangle .
\end{eqnarray}

\section*{Acknowledgements}

Work supported by the Spanish MEC through  FIS2009-13430-C02-C01
and Consolider NanoLight (CSD2007-00046) research grants, the former financing the work of JMA.

\end{document}